\documentclass[10pt, a4paper]{article}

\usepackage{jcappub}
\usepackage{graphicx}
\usepackage{amsmath,amssymb}
\usepackage{bm}

\def\be{\begin{equation}}
\def\ee{\end{equation}}
\def\ba{\begin{eqnarray}}
\def\ea{\end{eqnarray}}

\def\gdc{\gamma_{ D}}
\def\goc{\gamma_{\omega}}
\def\gbc{\gamma_{B}}
\def\GeV{\rm \, GeV}
\def\TeV{\rm \, TeV}
\def\eV{\rm \, eV}
\def\cm{\rm \, cm}
\def\vca{{\bf {\cal A}}}
\def\vce{{\bf {E}}}
\def\vcb{{\bf {B}}}
\def\vcy{{\bf {Y}}}

\def\vca{{\bf {A}}}
\def\vcz{{\bf {Z}}}
\def\ecp{\epsilon_{\text{\sc cp}}}
\newcommand{\per}{\, .}
\newcommand{\com}{\, ,}
\newcommand{\eref}[1]{Eq.~(\ref{#1})}

\newcommand{\eg}{{\it e.g.}}
\newcommand{\ie}{{\it i.e.}}

\usepackage{epsfig}     
\usepackage{color}              
\usepackage{graphicx}   
\usepackage{hyperref}
\hypersetup{
    colorlinks=true,
    linkcolor=red,
    citecolor=blue,
}

\begin{document}

\title{
Leptogenesis and Primordial Magnetic Fields
}

\date{\today}

\author[]{Andrew J. Long}
\author[]{, Eray Sabancilar} 
\author[]{, and Tanmay Vachaspati}
\affiliation[]{ 
Physics Department, Arizona State University, Tempe, Arizona 85287, USA.
}

\emailAdd{andrewjlong@asu.edu}
\emailAdd{eray.sabancilar@asu.edu}
\emailAdd{tvachasp@asu.edu}


\abstract{
The anomalous conversion of leptons into baryons during leptogenesis is shown
to produce a right-handed helical magnetic field; in contrast, the magnetic field
produced during electroweak baryogenesis is known to be left-handed. If the cosmological
medium is turbulent, the magnetic field evolves to have a present day coherence scale 
$\sim 10 \, {\rm pc}$ and field strength $\sim 10^{-18} \, {\rm Gauss}$.
This result is insensitive to the energy scale at which leptogenesis took place.
Observations of the amplitude, 
coherence scale, and helicity of the intergalactic magnetic field promise to provide 
a powerful probe of physics beyond the Standard Model and the very early universe.  
}

\maketitle


\section{Introduction}
\label{sec:intro}

Intergalactic magnetic fields may be relics from the early universe, and their detection could provide an important probe of early universe cosmology, high energy particle interactions that are not in reach of terrestrial experiments, and violations of fundamental CP symmetry in cosmology (for a recent review, see \cite{Durrer:2013pga}). 
Current upper bounds on the magnetic field strength derive from the anisotropy and polarization of the cosmic microwave background; these are at the level of $B \lesssim 10^{-9}~{\rm G}$ at the Mpc scale assuming a scale-invariant power spectrum \cite{Ade:2013zuv}.  
A lower bound, $B \gtrsim 10^{-16}~{\rm G}$, has also been derived from blazar data \cite{Neronov:1900zz, Tavecchio:2010mk}, although these claims are under further scrutiny and debate \cite{Tavecchio:2010mk, Neronov:2010bi, Dolag:2010ni, Broderick:2011av, Miniati:2012ge, Schlickeiser:2012xx}.
The hint of an all pervasive magnetic field in the universe provides compelling motivation to consider particle physics processes in the early universe that can generate such a field.  
In turn, the detection of a cosmological magnetic field together with its power spectrum and topology (helicity) would help to identify the epoch of magnetogenesis and thereby provide information about the state of the early universe. 
It has been shown, for instance, that a cosmological-scale magnetic field could be generated during electroweak baryogenesis and that its helicity density is directly proportional to the baryon asymmetry of the universe \cite{Cornwall:1997ms,Vachaspati:2001nb}.  
In this way, observations of the spectrum and helicity of cosmological magnetic fields would provide an additional handle on the origin of the baryon asymmetry of the universe.  

In this paper, we study the generation of a primordial magnetic field in conjunction with the creation of the baryon asymmetry of the universe through leptogenesis \cite{Fukugita:1986hr} (see \cite{Buchmuller:2005eh,Davidson:2008bu} for reviews).  
An overview of standard ``vanilla'' leptogenesis and the magnetogenesis scenario that we consider are described as follows.  
First, a lepton asymmetry is created when decays of the Majorana neutrino go out of equilibrium at a temperature $T = O(M_{N_R})$ where $M_{N_R}$ is the mass of the heavy Majorana neutrino.  
This mass scale is tied to the mass of the light neutrinos by the seesaw mechanism \cite{Fukugita:1986hr}, and typically $10^{9}~{\rm GeV} < M_{N_{R}} < 10^{15}~ {\rm GeV}$; as a fiducial scale we will use $M_{N_R} = 10^{10} \GeV$ for our numerical estimates.  
Second, the lepton asymmetry is converted into a baryon asymmetry via anomalous processes that violate the conservation of baryon- and lepton-numbers.  
These processes are sourced by extended gauge field configurations, and, as we discuss further below, it is therefore reasonable to expect these configurations to leave behind relic magnetic fields.  
Since leptogenesis occurs prior to electroweak symmetry breaking, the anomalous lepton-to-baryon number conversion can only produce a hypercharge magnetic field.  
The evolution of this field can be affected by the turbulence of the electroweak plasma and the chirality of this medium.  
When the electroweak phase transition occurs at $T \sim 10^{2} \GeV$, the hypercharge field is converted into an electromagnetic field.  
In this way, we argue that magnetogenesis is an inevitable consequence of leptogenesis.  

The anomalous $(B+L)$ current can be sourced by both non-Abelian and Abelian extended gauge field configurations \cite{tHooft:1976up}. 
For the non-Abelian case, these configurations are known as sphalerons at finite temperature \cite{Manton:1983nd,Klinkhamer:1984di}, and in the Standard Model there is one sphaleron associated with each of the strong and weak isospin gauge groups.  
The electroweak sphaleron, in particular, has been shown to produce a helical magnetic field as a ``by-product'' of its $L$-to-$B$ interconversion \cite{Vachaspati:2001nb,Copi:2008he,Chu:2011tx}.  
The Abelian field configuration is sometimes referred to as a hypermagnetic knot \cite{Giovannini:1999by,Jackiw:1999bd}, but we will adopt the terminology ``hypercharge sphaleron'' in order to mirror the non-Abelian case.  
The non-trivial topology of the Yang-Mills vacuum leads to a critical distinction between the two cases:  the non-Abelian configurations mediate ``vacuum-to-vacuum transitions'' in which the gauge field strength tensor vanishes asymptotically, but the Abelian configuration 
necessitates a non-vanishing field strength tensor at either early or late times.  
Consequently, the hypercharge sphaleron plays a key role in magnetogenesis from leptogenesis.  

The evolution of the hypermagnetic field from the time of magnetogenesis, $T \sim 10^{10} \GeV$, until the electroweak phase transition at $T \sim 10^{2} \GeV$, is governed by the magnetohydrodynamical (MHD) equations, which are coupled to the Boltzmann equations for the various chemical asymmetries.  
These equations encode a number of physical effects that, for the purposes of making order of magnitude estimates, can be disentangled and discussed separately in lieu of solving the equations directly.  
Namely, we will consider the evolution of the magnetic field subject to the Hubble expansion, viscous dissipation, chiral effects, and turbulence.  
We will see that the magnetic field correlation length at the point of production is large enough to evade immediate dissipation by the diffusive effect of the medium's viscosity.  
The fate of the field, then, is controlled by the dissipation due to Hubble expansion, unless some means of supporting the field strength is available.  
We first note that the medium acquires a chiral asymmetry during leptogenesis, and it is known that the chiral magnetic and vortical effects \cite{Vilenkin:1980fu, Vilenkin:1979ui} can lead to an amplification of the field strength \cite{Joyce:1997uy,Boyarsky:2011uy,Tashiro:2012mf}.  
We will see, however, that the magnitude of the chiral asymmetry, which is related to the observed baryon asymmetry, is too small for the chiral effects to be operative.  
As a second possibility, we suppose that the medium is turbulent at the time of magnetogenesis.  
In the presence of turbulence small-scale helical magnetic fields will experience a so called ``inverse cascade'' as the coherence scale grows to astrophysically interesting length scales \cite{Jedamzik:1996wp, Banerjee:2004df, Kahniashvili:2012uj, Saveliev:2013uva}
Turbulence may arise, for example, as the result of bubble percolation at a first order phase transition, however it should be emphasized that leptogenesis does not require a first order phase transition, and the presence of turbulence has to be an additional assumption.  

We start by describing the leptogenesis model in Sec.~\ref{sec:lepto}.  
We follow the lepton number equilibration process in Sec.~\ref{sec:equil} and the magnetic field production in Sec.~\ref{sec:magnetogenesis}.  
In Sections~\ref{sec:evol}~and~\ref{sec:turbulence}, we discuss the evolution of the magnetic field for the case of a non-turbulent and turbulent medium, respectively.  
We also estimate the magnetic field coherence scale and field strength at the present epoch.  
We conclude in Sec.~\ref{sec:disc}.  
The results critically depend on a number of signs in the analysis, and we have found several errors in the literature. 
To clarify some of these signs, we discuss our conventions in Appendix~\ref{sec:signs}.
We summarize the anomaly equations in Appendix~\ref{sec:anomeqs} for convenience.


\section{Leptogenesis}
\label{sec:lepto}

In this section and the next, we review the key features of standard Majorana neutrino leptogenesis \cite{Fukugita:1986hr} (see, \eg, \cite{Buchmuller:2005eh,Davidson:2008bu}).  
The lightest Majorana neutrino decays through its Yukawa interaction with the Higgs and lepton doublets, 
\begin{align}
	\mathcal{L}_{\rm int} = -(\lambda^{\nu})^{ij} \epsilon^{ab} (L^{\dagger})_{a , i} 
	(\Phi^{\ast})_{b} (N_R)_j + {\rm h.c.} \com
\end{align}
where $N_R$, $L = (\nu_L \, , \, e_L)$, and $\Phi = (\Phi^+ \, , \, \Phi^0)$ are the Majorana neutrino, lepton doublet, and Higgs doublet fields, respectively, and $i,j$ label generations.  
Specifically, half of the decay channels contain leptons in the final state, $N_R \to (e_L)_i \Phi^+, (\nu_L)_i \Phi^0$, and half contain anti-leptons, $N_R \to (e_L)^c_i \Phi^-, (\nu_L)^c_i (\Phi^0)^c$, where we use the superscript ``C'' to indicate the anti-particle.  
The rate of decays into leptons and anti-leptons, denoted $\Gamma_L$ and $\Gamma_{L^c}$ respectively, are biased due to CP-violating phases in the Yukawa matrix.  
Thus the CP-violation parameter \cite{Buchmuller:2005eh},
\begin{align}\label{eq:cpv_def}
	\ecp \equiv \frac{\Gamma_L - \Gamma_{L^c}}{\Gamma_L + \Gamma_{L^c}} \com
\end{align}
can be estimated as 
\begin{align}\label{eq:cpv}
	\ecp \approx \left| \lambda^{\nu} \right|^2 \sin \theta_{\rm CP}~,
\end{align}
where $\theta_{\rm CP}$ is the CP-violating phase, and for $M_{N_R} = 10^{10} \GeV$ the Yukawa coupling is $\left| \lambda^{\nu} \right|^2 = O(10^{-6})$.  

After decay of the Majorana neutrino goes out of equilibrium, the medium contains a chiral and a lepton asymmetry.  
It is helpful to introduce chemical potentials to keep track of these asymmetries.  
Let $n_X(t)$ be the spatially averaged number density of the species labeled $X$, and let $n_{X^c}(t)$ be the number density of its charge conjugate.  
The particle number asymmetry is related to the chemical potential $\mu_X(t)$ by
\begin{equation}\label{nX-nXc}
	n_{X} - n_{X^c} \approx \frac{g T^3}{6}  
	\begin{cases} \frac{\mu_X}{T} + O\left( (\mu_X / T)^3 \right) & {\rm fermions}~, \\ 2 
	\frac{\mu_X}{T} + O\left( (\mu_X / T)^3 \right) & {\rm bosons}~, 
	\end{cases}
\end{equation}
where $g$ is the number of dynamical degrees of freedom ({\it e.g.}, isospin, color) associated with the particle labeled $X$.  
There is a chemical potential for each of the Standard Model particle species:  the left-handed quark doublet $Q^i = (u_L^i \, , \, d_L^i)$, the right-handed quark singlets $u_R^i$ and $d_R^i$, the left-handed lepton doublet $L^i = (\nu_L^i \, , \, e_L^i)$, the right-handed electron singlet $e_R^i$, and the scalar Higgs doublet $\Phi = (\Phi^+ \, , \, \Phi^0)$.  
The index $i$ indicates the generation and takes values $i = 1, 2, 3$.  
At all temperatures of interest, the gauge interactions are in equilibrium, and they enforce an equality of the asymmetries carried by different components of a given multiplet.  
For example, $\mu_{u_R^i , {\rm red}} = \mu_{u_R^i , {\rm blue}} = \mu_{u_R^i , {\rm green}} \equiv \mu_{u_R^i}$ or $\mu_{\Phi^+} = \mu_{\Phi^0} \equiv \mu_{\Phi}$.  

Decay of the Majorana neutrinos provides the initial asymmetries carried by each of the Standard Model species.  
For simplicity, we will assume that decays of $N_R$ are uniform into the three generations.  
If $n_{N_R} \sim T^3$ is the number density of Majorana neutrinos before they decay, then in the instantaneous decay approximation, the CP asymmetry provides the initial condition 
\begin{equation}\label{eq:initial_condit}
	\mu_{L^i}(t_0) = \mu_0~, 
	\qquad 
	2\mu_{\Phi}(t_0) = 3 \mu_0~,
	 \qquad
	\mu_{Q^i}(t_0) = \mu_{u_R^i}(t_0) = \mu_{d_R^i}(t_0) = \mu_{e_R^i}(t_0) =0 \com
\end{equation}
where
\begin{align}\label{eq:mu0_def}
	\frac{\mu_0}{T} \approx f_{wo} \ecp \frac{n_{N_R}}{T^3} \per
\end{align}
The washout factor, $f_{wo} < 1$, accounts for a suppression of the asymmetry due to inverse decays and lepton-number-violating re-scatterings.  
We will assume $f_{wo} = O(1)$, and therefore the estimate in \eref{eq:cpv} yields
\begin{align}\label{eq:mu0_estimate}
	\frac{\bigl| \mu_0 \bigr|}{T} \lesssim O(10^{-6}) \per
\end{align}
Note that the nonzero $\mu_{L^i}$ represents both a chiral and a lepton asymmetry.  
The factor of $3$ in $\mu_{\Phi}(t_0)$ appears because we assume that $N_R$ decays equally into each generation of $L^i$, and the factor of $2$ multiplying $\mu_{\Phi}$ appears as a consequence of spin statistics [see Eq.~(\ref{nX-nXc})].  


 \section{Equilibration}
 \label{sec:equil}
 
The initial asymmetries, given by \eref{eq:initial_condit}, are subsequently converted into a baryon asymmetry as the system returns to chemical equilibrium.  
There are various perturbative and non-perturbative particle physics reactions that participate in this equilibration.  
These reactions, which are summarized in Table~\ref{tab:processes}, are the Yukawa interactions, the strong sphaleron, the weak sphaleron, and the ``hypercharge sphaleron."  
The only other perturbative interactions in the Standard Model are the gauge interactions and the Higgs self-interaction but, since these do not change fermion number, they are not relevant to the re-distribution of the various asymmetries.  
As we discuss these reactions below, we are particularly interested in their rates relative to the Hubble expansion rate 
\be\label{H}
H =\sqrt{\frac{8 \pi}{3} \frac{g_{\ast} \pi^2}{30}} \, \frac{T^2}{ M_P} \approx 140\left(\frac{T}{10^{10}~{\rm GeV}}\right)^2~{\rm GeV}~,
\ee
where $M_P = 1.22\times 10^{19} \GeV$ is the Planck mass and $g_{\ast} = 106.75$ is the number of relativistic species at $T \gtrsim 1 \TeV$.  

The Standard Model admits three types of Yukawa interactions, up-type, down-type, and electron-type, with rates $\Gamma_{\rm Yuk} \sim h^2 T / 8 \pi$ where $h$ is the appropriate Yukawa coupling.
The Yukawa interactions violate chirality (and Higgs number) and tend to relax the chiral asymmetries to zero.  
The first (lightest) generation Yukawa interactions only come into equilibrium ($\Gamma_{\rm Yuk} < H$) below the temperature \cite{Campbell:1992jd}
\begin{align}\label{eq:Tyuk}
	T_{\rm Yuk} \approx 10^2-10^4 \TeV \com
\end{align}
whereas the second and third generation interactions are already in equilibrium above this temperature.  
Although leptogenesis occurs at $T \lesssim 10^{10}~{\rm GeV}$, we will assume for simplicity\footnote{Further assumptions about the couplings of $N_R$ to each of the generations would be necessary as well if we were to treat all three generations.} that all of the Yukawa interactions are out of equilibrium down to $T_{\rm Yuk}$.  
We expect that this assumption will introduce an $O(1)$ error in our calculation, but will not change the sign nor the order of magnitude of the resultant asymmetries.  
If the Yukawa interactions are ignored, the Higgs particle number is also conserved, and we have
\begin{equation}\label{Deltamu}
	\Delta \mu_\Phi = 0~.
\end{equation}
Additionally, without the Yukawa interactions there is no distinction between the three generations, and we can drop the generation index on the various chemical potentials.

The three sphaleron reactions are particularly important for the success of leptogenesis, because they violate the conservation of baryon-plus-lepton-number, which derives from an anomalous symmetry.  
A complete list of chiral anomaly equations is given in Appendix~\ref{sec:anomeqs}, where it may be seen that the anomalous currents are sourced by gauge field configurations with nonzero Chern-Simons number.  
The sphaleron reactions are therefore mediated by non-perturbative, extended gauge field configurations.  
In the ${\rm SU}(3)_{c}$ sector the strong sphaleron mediates a vacuum-to-vacuum transition which induces a chirality-violating process among the colored particles \cite{McLerran:1990de}.  
We say that the transition is ``vacuum-to-vacuum'' because the ${\rm SU}(3)_c$ gauge field strength tensor vanishes asymptotically.  
The spatial and temporal extent of the strong sphaleron are set by the scale $(\alpha_{\rm s} T)^{-1}$ where $\alpha_{\rm s} \equiv g_{\rm s}^2 / 4 \pi \sim 0.1$ is the strong fine structure constant\footnote{Here and below we neglect the running of the coupling constants between the scale at which they are measured and the scale of leptogenesis.  This approximation is consistent with our order of magnitude estimates.  }.  
The rate per unit volume for this reaction is then naively estimated as $(\alpha_{\rm s} T)^{4}$, but an additional factor of $\alpha_{\rm s}$ is required to account for the damping of non-Abelian gauge fields in the plasma \cite{Arnold:1996dy}.  
An additional dimensionless prefactor is determined from lattice simulations wherein the rate is found to be \cite{Moore:1997im}
\be\label{strongsph}
\Gamma_{\rm s} \sim 100 \alpha_{\rm s}^5 T \per
\ee

Similarly in the ${\rm SU}(2)_{L}$ sector, the weak sphaleron mediates a vacuum-to-vacuum transition \cite{Manton:1983nd,Klinkhamer:1984di}, and induces a reaction among the weakly interacting particles at a rate \cite{Moore:1997sn, D'Onofrio:2012jk}
\be\label{weaksph} 
\Gamma_{\rm w} \sim 25\alpha_{\rm w}^5 T~, 
\ee
where $\alpha_{\rm w} \equiv g^2 / 4 \pi \sim 1/30$ is the weak fine structure constant.  
The weak sphaleron plays a critical role in leptogenesis since it violates baryon-number and lepton-number each equally \cite{Kuzmin:1985mm}, and thereby converts a negative lepton-number into a positive baryon-number.

\begin{table}[t]
\begin{center}
\begin{tabular}{|c|c|c|c|}
\hline
Name & Particle Reaction & Rate & Chemical Equilibrium \\
\hline
$\begin{array}{c}\text{Up-Type} \\ \text{Yukawa} \end{array}$ 
& $\begin{array}{l} d_L^i + \Phi^+ \leftrightarrow u_R^i \\ u_L^i + \Phi^0 \leftrightarrow u_R^i \end{array}$ 
& $\frac{h_{u^i}^2}{8 \pi} T$ 
& $\mu_{Q^i} + \mu_{\Phi} - \mu_{u_R^i} = 0$ \\
\hline
$\begin{array}{c}\text{Down-Type} \\ \text{Yukawa} \end{array}$ 
& $\begin{array}{l} u_L^i \leftrightarrow \Phi^+ + d_R^i \\ d_L^i \leftrightarrow \Phi^0 + d_R^i \end{array}$ 
& $\frac{h_{d^i}^2}{8 \pi} T$ 
& $\mu_{Q^i} - \mu_{\Phi} - \mu_{d_R^i} = 0$ \\
\hline
$\begin{array}{c}\text{Electron-Type} \\ \text{Yukawa} \end{array}$ 
& $\begin{array}{l} \nu_L^i \leftrightarrow \Phi^+ + e_R^i \\ e_L^i \leftrightarrow \Phi^0 + e_R^i \end{array}$ & $\frac{h_{e^i}^2}{8 \pi} T$ 
& $\mu_{L^i} - \mu_{\Phi} - \mu_{e_R^i} = 0$ \\
\hline 
$\begin{array}{c}\text{Strong} \\ \text{Sphaleron} \end{array}$ 
& $\sum_i \left( u_L^i + d_L^i \right) \leftrightarrow \sum_i \left( u_R^i + d_R^i \right)$ 
& $100\alpha_{\rm s}^5 T$ 
& $\sum_i \left( 2 \mu_{Q^i} - \mu_{u_R^i} - \mu_{d_R^i} \right) = 0$ \\
\hline
$\begin{array}{c}\text{Weak} \\ \text{Sphaleron} \end{array}$ 
& $\begin{array}{l}\sum_i \left( u_L^i + d_L^i + d_L^i + \nu_L^i \right) \leftrightarrow 0 \\ \sum_i \left( u_L^i + u_L^i + d_L^i + e_L^i \right) \leftrightarrow 0 \end{array}$ 
& $25\alpha_{\rm w}^5 T$ 
& $\sum_i \left( 3\mu_{Q^i} + \mu_{L^i} \right) = 0$ \\
\hline
$\begin{array}{c}\text{Hypercharge} \\ \text{Sphaleron} \end{array}$ 
& $\begin{array}{l} \sum_i \left(9 (\nu_L^i + e_L^i) + \sum_{c} (u_L^i + d_L^i) \right) \\ \qquad \longleftrightarrow \sum_i \left( \sum_c (16 u_R^i + 4 d_R^i) + 36 e_R^i \right) \end{array}$ 
& $\alpha_Y^4 T$ 
& $\begin{array}{l} \sum_i \Bigl( 18 \mu_{L^i} + 6 \mu_{Q^i} - 48 \mu_{u_R^i} \\ \qquad - 12 \mu_{d_R^i} - 36 \mu_{e_R^i} \Bigr) = 0 \end{array}$ \\
\hline
\end{tabular}
\end{center}
\caption{\label{tab:processes}
The various perturbative and non-perturbative reactions discussed in the text.  Summation over color
indicated by $\sum_c$; summation over generations is denoted by $\sum_i$. 
}
\end{table}

In the ${\rm U}(1)_Y$ hypercharge sector a third anomalous reaction, associated with the Abelian anomaly [see Eq.~(\ref{eq:anomalyequations})], is mediated by a hypercharge gauge field configuration that we will call the ``hypercharge sphaleron.''  
As above, we estimate the spatial and temporal extent of this field configuration as $(\alpha_Y T)^{-1}$ where $\alpha_Y \equiv g'^{\,2} / 4 \pi \sim 0.01$ is the hypercharge fine structure constant.  
We can similarly estimate the rate, but since the field is Abelian the magnetic component is not damped and the additional factor of $\alpha_{Y}$ is not required.  
Thus we take the rate to be
\be\label{hypsph}
\Gamma_{Y} \sim \alpha_Y^4 T \per
\ee
Note that all three sphaleron processes are faster than the Hubble rate, given by Eq.~(\ref{H}), at $T \sim 10^{10} \GeV$ hence they are all in thermal equilibrium during the leptogenesis epoch.  

As we will see below, the initial correlation length of the hypermagnetic field is tied to the length scale of the hypercharge sphaleron, $l_Y$.  
Because of the significance of $l_Y$ for our magnetogenesis calculation, we provide here a heuristic derivation of the relationship $l_Y \sim (\alpha_Y T)^{-1}$.  
(Here we parallel the discussion of \cite{Arnold:1996dy}).  
We are interested in field configurations with a nonzero Chern-Simons number $\Delta N_{\rm CS}^{(Y)} \sim \alpha_Y \int d^4 x Y_{\mu \nu} \tilde{Y}^{\mu \nu} \sim \alpha_Y R^4 B^2$ where $R$ is the typical length scale of the configuration, and $B$ is the typical field strength.  
Requiring $\Delta N_{\rm CS}^{(Y)} = O(1)$ gives $B \sim 1 / (\sqrt{\alpha_Y} R^2)$.  
The energy of such a configuration is $E \sim B^2 R^3 \sim 1 / (\alpha_Y R)$.  
At a given temperature $T$ the smallest configuration that is not Boltzmann suppressed has a size given by $E \sim T$ or $R \sim 1 / (\alpha_Y T)$; large $R$ configurations are entropically suppressed.
Then it follows that the rate per unit volume of these transitions is $\Gamma / V \sim 1 / R^4 \sim (\alpha_Y T)^{4}$, and the rate in an inter-particle 
 volume is $\Gamma \sim \alpha_Y^4 T$.

An anomalous reaction associated with the ${\rm U}(1)_Y$ hypercharge sector is not usually included in leptogenesis calculations.  
This may be related to the fact that the hypercharge sphaleron, unlike the strong and weak sphalerons, is not a vacuum-to-vacuum transition.  
This is a consequence of the trivial topology of the vacuum manifold of the Abelian gauge theory, and for this case the Chern-Simons number is proportional to the field strength tensor \cite{Cheng:1984}.  
As a result, if the Abelian Chern-Simons number changes in a process, then either the initial or the final state (or both) cannot be the vacuum.  
This exchange of energy between the particle sector and the gauge sector provides the basis of magnetogenesis.  
As we will see, the energy exchange is minimal, perhaps justifying the neglect of the hypercharge sphaleron in leptogenesis calculations.  
However, we should emphasize that field configurations with nonzero Chern-Simons number are known to exist, \eg, hypermagnetic knots \cite{Giovannini:1999by,Jackiw:1999bd} and linked magnetic flux tubes \cite{Vachaspati:2001nb}.  
We use the term ``hypercharge sphaleron'' to refer to any ${\rm U}(1)_Y$ field configuration that interpolates between a vacuum configuration and a configuration with non-zero Chern-Simons number, which in MHD corresponds to magnetic field ``helicity'' (see below).  
Such a field configuration will source the anomaly equation for each Standard Model fermion field since they all carry hypercharge. Particle production due to the anomaly 
can also be understood as the emergence of filled states from
the Dirac sea into the positive energy sector, just as in the case of the Abelian
chiral anomaly in 3+1 dimensions
\cite{Adam:2000gt}. 

The hypercharge sphaleron carries an Abelian Chern-Simons number that sources the anomaly equations, Eq.~(\ref{eq:anomalyequations}), and induces anomalous particle production.   
The corresponding particle reaction can be deduced by noting that the divergence of each Standard Model field current is proportional to the square of that field's hypercharge, and that the divergences of left-chiral and right-chiral fields are opposite in sign.  
It follows that the hypercharge sphaleron induces the reaction
\begin{align}\label{eq:hypercharge_sphaleron}
	\sum_i \left(9 (\nu_L^i + e_L^i) + \sum_{c} (u_L^i + d_L^i) \right) \longleftrightarrow \sum_i \left( \sum_c (16 u_R^i + 4 d_R^i) + 36 e_R^i \right)~,
\end{align}
where $\sum_{i}$ is a sum over generations and $\sum_{c}$ is a sum over colors. Note that the factors $9, 1, 16, 4,$ and $36$ are proportional to the square of the corresponding field's hypercharge.  
Both the left and right sides of this reaction are color and weak isospin singlets.  
Since a nonzero Abelian Chern-Simons number also sources the baryon and lepton number currents, see \eref{baryonanomaly}, it is not surprising that the reaction in \eref{eq:hypercharge_sphaleron} violates these charges as well.  
This reaction involves $120 N_g=360$ separate particles where $N_g = 3$ is the number of generations.  
If any particles were removed, this reaction could not be made to conserve the gauge charges without reducing to one of the other reactions already discussed.  
When the reaction in \eref{eq:hypercharge_sphaleron} is in chemical equilibrium, the chemical potentials satisfy 
\begin{align}\label{eq:HS_chempot}
	\sum_i \left( 18 \mu_{L^i} + 6 \mu_{Q^i} - 48 \mu_{u_R^i} - 12 \mu_{d_R^i} - 36 \mu_{e_R^i} \right) = 0 \per
\end{align}

One can easily verify by inspecting Table~\ref{tab:processes} 
(see also Appendix~\ref{sec:anomeqs}) that each of the reactions respects the 
conservation of hypercharge and ``$B-L$'' (baryon minus lepton number).  
Consequently, the corresponding asymmetries are conserved
\begin{align}\label{eq:Consv_Law_1}
	\Delta \mu_Y = 0~,
	\qquad
	\Delta (\mu_{\rm bar} - \mu_{\rm lep} ) = 0 \com
\end{align}
where
\begin{align}
\mu_{Y} &\equiv \sum_i \left( 6 y_Q \mu_{Q^i} + 3 y_{u_R} \mu_{u_R^i} + 3 y_{d_R} \mu_{d_R^i} + 2 y_{L} \mu_{L^i} + y_{e_R} \mu_{e_R^i} \right) + 2 y_{\Phi} (2\mu_{\Phi})~, \\
	\mu_{\rm lep} &\equiv \sum_{i} \left( 2\mu_{L^i} + \mu_{e_R^i} \right)~, \\
	\mu_{\rm bar} &\equiv \sum_i \frac{1}{3} \left( 6\mu_{Q^i} + 3\mu_{u_R^i} + 3\mu_{d_R^i} \right) \label{def_mubar}~,
\end{align}
and $y_Q=1/3$, $y_{u_R}= 4/3$, $y_{d_R}=-2/3$, $y_L=-1$, $y_{e_R}=-2$, and $y_{\Phi} = 1$ are the hypercharge assignments. 
The various prefactors are the multiplicites of each multiplet; for example, 
$2 \mu_{L^i} = \mu_{\nu_L^i} + \mu_{e_L^i}$.  

The right-most column of Table~\ref{tab:processes} gives the equilibrium conditions on the chemical potentials.  
Assuming all the Yukawa interactions are slow and out of equilibrium at temperatures above $T_{\rm Yuk} \gtrsim 10^{2} \TeV$, see \eref{eq:Tyuk}, then only the three sphaleron processes in Table~\ref{tab:processes} are operative.  
Using the conservation laws, Eqs.~(\ref{Deltamu})~and~(\ref{eq:Consv_Law_1}), we obtain the chemical potentials for different particle species as 
\begin{equation}\label{musolution}
	\mu_{Q} = -\frac{7}{44}\mu_0~, \ \ 
	\mu_L = \frac{21}{44} \mu_0~, \ \ 
	\mu_{u_R} = -\frac{\mu_0}{11}~, \ \
	\mu_{d_R} = -\frac{5}{22} \mu_0~, \ \
	\mu_{e_R} = \frac{9}{22} \mu_0~, \ \ 
	\mu_{\Phi} = \frac{3\mu_0}{2}~.
\end{equation}
Using Eq.~(\ref{def_mubar}) we obtain the baryon chemical potential,
\begin{equation}
	\mu_{\rm bar} = - \frac{7}{11} \mu_0~.
	\label{mubar}
\end{equation}
The observed baryon asymmetry is $n_{\rm bar}/s \approx 1 \times 10^{-10}$ \cite{Ade:2013zuv} where $s = (2 \pi^2 / 45) \, g_{\ast} T^3$ is the entropy density.  
Using Eq.~(\ref{nX-nXc}) the number density of baryon number can be expressed as $n_{\rm bar} = \mu_{\rm bar} T^2/6$. Thus, in order to obtain the observed baryon asymmetry, the initial lepton asymmetry must be
\begin{equation}\label{lepbar}
	\frac{\mu_0}{T} 
	= -\frac{11}{7} \frac{\mu_{\rm bar}}{T} 
	= -\frac{66}{7} \frac{2\pi^2}{45} g_{\ast} \frac{n_{\rm bar}}{s} 
	\approx - 5 \times 10^{-8}~.
\end{equation}
Note that $\mu_0 < 0$ as we expected since $B-L$ is conserved.  
Moreover, we find that the required value of $\mu_0$ is consistent with our previous bound $\mu_0 / T \lesssim O(10^{-6})$, given by \eref{eq:mu0_estimate}.  
Using \eref{eq:mu0_def} with the assumptions $f_{wo} \lesssim O(1)$ and $n_{N_R} = O(T^3)$, we find that 
\begin{align}\label{eq:ecp_numerical}
	\ecp \gtrsim O(10^{-8})
\end{align}
is required in order for leptogenesis to match the observed baryon asymmetry.  
Alternatively, if the Majorana neutrino decay occurs far from equilibrium, $n_{N_R} \gg O(T^3)$, then smaller values of $\ecp$ are allowed.  

In deriving \eref{musolution} we have based our calculation on equilibrium conditions and assumptions about the relative rates of the various reactions.  
We believe that this approach is reliable for making order of magnitude estimates.  
One can perform a more detailed analysis of the equilibration process by solving the kinetic equations associated with the chemical potentials, and moreover the evolution of magnetic fields can be studied via the coupling of the chiral asymmetries to the MHD equations and the gauge field kinetic equations.  
Such an approach was recently pursued in Refs.~\cite{Giovannini:1997eg,Semikoz:2012ka,Dvornikov:2012rk,Semikoz:2013xkc}, but taking different initial conditions than \eref{eq:initial_condit}, neglecting the hypercharge sphaleron reaction, and neglecting turbulence.  
In the present paper we simply work with the equilibrium solution described above and ignore the dynamics that leads to the equilibrium state.


\section{Magnetogenesis}
\label{sec:magnetogenesis}

As we discussed above, the hypercharge sphaleron does not mediate a vacuum-to-vacuum transition, but instead it must be accompanied by the production of a hypercharge magnetic field.
The spatially averaged helicity density of this field is defined as 
\begin{align}\label{eq:hY_def}
	h_Y(t) \equiv \lim_{V \to \infty} \frac{1}{V} \int_V d^3x ~ \vcy \cdot \vcb_Y~,
\end{align}
where ${\bf Y}$ is the vector potential and ${\bf B}_{Y} = {\bm \nabla} \times {\bf Y}$ is the hypermagnetic field.  
We calculate $h_{Y}$ by integrating the anomaly equation [see \eref{eq:anomalyequations}], 
\begin{equation}\label{eq:anomaly_equation}
	\partial_{\mu} j^{\mu}_{e_R} = - \frac{g'^{\,2}}{16 \pi^2} Y_{\mu \nu} \tilde{Y}^{\mu \nu}~,
\end{equation}
over the spacetime volume as in \eref{eq:Deltah}.  
Then using 
\begin{align}
	n_{e_R} - n_{e_R^c} = \lim_{V \to \infty} \frac{1}{V} \int_{V} d^3 x \, j_{e_R}^0 \com
\end{align}
along with the initial condition $h_Y(-\infty) = (n_{e_R} - n_{e_R^c})\bigr|_{-\infty} = 0$, we obtain 
\begin{equation}
	h_Y = - \frac{2\pi}{\alpha_Y} \left( n_{e_R} - n_{e_R^c} \right) \per
\end{equation}
To relate this to the baryon asymmetry, $n_{\rm bar} / s \sim 10^{-10}$, we use Eqs.~(\ref{nX-nXc}), (\ref{musolution}) and (\ref{mubar}) to find 
 \begin{align}\label{eq:hY_numerical}
	\frac{h_Y}{T^3} = \frac{2 \pi^2 g_{\ast}}{35 \alpha_Y} \frac{n_{\rm bar}}{s} \sim 10^{-6} \per
\end{align}
This equation is a central result of our paper.  
It indicates the connection between the baryon asymmetry created by leptogenesis and the helicity of the magnetic field created during the concurrent process of magnetogenesis.  
We find that the magnitude of the helicity density is set by the size of the baryon asymmetry of the universe.  
Moreover, since $h_Y > 0$ we find that the hypercharge magnetic field produced in this way carries right-handed helicity. 

At the level of this analysis, we do not calculate the full spectrum of the hypermagnetic field, but instead we characterize it using the typical correlation length, $\xi$, and field strength, $B_Y$, which is related to the magnetic field energy density $\rho_B \sim B_Y^2$.  
The correlation length is controlled by the size of the hypercharge sphaleron, and we expect
\begin{align}\label{eq:xi_initial}
	\xi \sim l_Y \sim (\alpha_Y T)^{-1}
\end{align}
at the time of magnetogenesis.  
The correlation length subsequently evolves as described in the next section.  

Using $\xi$ and $h_Y$, dimensional analysis allows one to estimate the magnetic field strength as $B_Y \sim \sqrt{h_Y / \xi} \sim 10^{-4} T^2$ and the energy density as $\rho_B \sim B_Y^2 \sim 10^{-8} T^4$.  
However, this result may be a significant underestimate since it is based on just the helicity alone.  
The enhancement can be understood with the following heuristic argument\footnote{A similar argument can be found in Ref.~\cite{Copi:2008he} in the context of magnetogenesis from electroweak baryogenesis.}.
Suppose that the decay of $N_R$'s in a certain volume produces $X+\delta X$ anti-leptons ($L^c$) and $X$ leptons ($L$) with the slight excess of $L^c$'s due to CP violation in the model. 
(These $L$ and $L^c$ are in excess of the already present thermal background.)
The estimate above that yielded $\rho_B \sim 10^{-8} T^4$ only ``counts'' the magnetic field produced due to the conversion of $\delta X$ $L^c$'s into baryons.  
Such an estimate is valid if the $X$ $L^c$'s annihilate instantly with the $X$ $L$'s, but more realistically, a fraction of the $X$ $L^c$'s and $X$ $L$'s will also convert into baryon number. 
This fraction is given by the ratio of the conversion rate, $\Gamma_Y \sim \alpha_Y^4 T$, to the annihilation rate, $\Gamma_{\rm ann} \sim \alpha_{\rm w}^2 T$, which proceeds predominantly through the weak interaction.  
Then the fraction is $\Gamma_Y/\Gamma_{\rm ann} \sim \alpha_Y^4/ \alpha_{\rm w}^2 \sim 10^{-5}$.
Since $\delta X$ is fixed by the observed baryon number density, $X$ is larger than $\delta X$ in inverse proportion to the CP violation factor in the model, $\ecp$ given by Eq.~(\ref{eq:cpv}).
Then the resulting energy density in the magnetic field may be significantly amplified by a factor $f \sim \alpha_Y^4/ (\ecp~ \alpha_{\rm w}^2)$, which is approximately $f \sim 10^3$ for $\ecp \sim 10^{-8}$.  

Based on dimensional analysis and the arguments of the preceding paragraph, we estimate the magnetic field energy density as 
\begin{equation}\label{hrhoB}
	\rho_B \sim \frac{\alpha_Y^4}{\ecp \alpha_{\rm w}^2} \frac{h_Y}{\xi} \per
\end{equation}
Using Eqs.~(\ref{eq:hY_numerical})~and~(\ref{eq:xi_initial}) we obtain
\begin{equation}\label{rhoB}
	\rho_B 
	\sim 10^{-5} \left(\frac{10^{-8}}{\ecp}\right) \, T^4 \com
\end{equation}
where we have normalized to the value of $\ecp$ given by \eref{eq:ecp_numerical}.  
This estimate shows that the backreaction of the magnetic field on the plasma can only be ignored for $\ecp \gtrsim 10^{-13}$.


\section{Magnetic Field Evolution in a Non-Turbulent Medium}
\label{sec:evol}

We now turn our attention to the evolution of the magnetic field in the post-leptogenesis era.  
As the initial condition, we assume that the magnetic field spectrum is peaked at a scale $\xi$ given by \eref{eq:xi_initial} with an energy density $\rho_B$ given by \eref{rhoB} and a helicity density given by \eref{eq:hY_numerical}.  
In lieu of solving the fully coupled MHD and Boltzmann equations, we will identify the relevant physical processes and address each of these in turn.  

As a first case, we suppose that the medium is not turbulent and that chiral effects can be neglected.  
Then the magnetic field will simply evolve subject to the diffusive equation\footnote{\label{fn:conventions}We work in rescaled comoving coordinates.  Any dimensionful quantities are rendered dimensionless by introducing factors of temperature.  For instance, the dimensionless comoving wavenumber satisfies $k_{\rm comoving} = k_{\rm physical} / T$.  The dimensionless conformal time, $\eta$, satisfies $d \eta = T(t) dt$ and is directly related to the temperature via $\eta = \sqrt{90 / (8 \pi^3 g_{\ast}) } M_P / T$.  }
\begin{equation}\label{diffmhd}
	\partial _\eta {\bm B}_Y  =  \gdc \nabla ^2 {\bm B}_Y~, 
\end{equation}
where the dimensionless parameter $\gamma_D$ is the diffusion coefficient, and it is given by $\gamma_D = 1 / \sigma$ where $\sigma \sim 100$ is the dimensionless electrical conductivity in the electroweak symmetric phase \cite{Turner:1987bw, Baym:1997gq, Arnold:2000dr}.  
The solution to \eref{diffmhd} is an exponential decay,
\begin{align}\label{eq:exp_decay}
	B(\eta, k) = B_{0}(k) \ {\rm exp}[- \gamma_D k^2 \eta] \per
\end{align}
The small scale modes (high $k$) are rapidly damped, whereas the larger modes (low $k$) are ``frozen'' until a time $\eta \sim 1 / \gamma_D k^2$.
For instance, in order for a mode to survive until recombination, $\eta_{\rm rec} \approx 10^{28}$, it must have comoving wavenumber $k \lesssim k_{\rm frozen}$ with
\begin{equation}\label{eq:kfrozen}
	k_{\rm frozen} = \frac{1}{\sqrt{\gdc \eta_{\rm rec}}} \sim 10^{-13} \per
\end{equation}
For a mode to survive until today, it must have an even smaller wavenumber.  
To assess the impact of diffusion on the magnetic field created during leptogenesis, we estimate $k = 1 / (\xi T)$ and take $\xi \sim (\alpha_Y T)^{-1}$ as in \eref{eq:xi_initial}.  
We obtain $k \sim \alpha_Y \sim 10^{-2}$ independent of the temperature of leptogenesis.  
Since $k \gg k_{\rm frozen}$ we see that the magnetic field will be rapidly damped out.  

As a second scenario, we will suppose that the medium carries a chiral asymmetry.  
This assumption is well-justified because, as we showed in Sec.~\ref{sec:equil}, the medium acquires a chiral asymmetry during leptogenesis.  
The generation and evolution of the magnetic field and chirality have been studied by several authors \cite{Joyce:1997uy,Giovannini:1997eg,Boyarsky:2011uy,Tashiro:2012mf,Boyarsky:2012ex,Semikoz:2012ka,Dvornikov:2012rk,Semikoz:2013xkc} in different contexts.  
In these analyses, it is customary to assume that the medium contains a large chiral asymmetry stored in the right-chiral electrons, $\mu_{e_R} / T \gtrsim 10^{-6}$, which is protected from chirality-flipping Yukawa interactions until $T \sim 10^2 \TeV$ \cite{Joyce:1997uy}.  
However, since leptogenesis must reproduce the observed baryon asymmetry, we have seen that the magnitude of the chiral asymmetry is typically $\mu / T \sim 10^{-8}$ (recall \eref{lepbar}), and therefore too small for the chiral effects to play any significant role.  
In the remainder of this section, we will make this point explicit.  

In the presence of a chiral medium with a velocity field ${\bf v}(x)$, the hypermagnetic field evolution is governed by the Abelian MHD equation \cite{Tashiro:2012mf}
\begin{equation}\label{mhdeq}
	 \partial _\eta {\bm B}_Y  = \nabla  \times ({\bm v} \times {\bm B}_Y)
	+ \gdc \nabla ^2 {\bm B}_Y 
	+ \gbc \nabla \times {\bm B}_Y 
	+ \goc \nabla \times (\nabla \times {\bm v}) \per
\end{equation}
The terms with the dimensionless coefficients $\gamma_{B}$ and $\gamma_{\omega}$ represent the chiral magnetic and chiral vortical effects, respectively.  
These coefficients are given by\footnote{There are two compensating sign errors in \cite{Tashiro:2012mf}: there should be a minus sign in their Eq.~(4), and also in their Eq.~(46). As a result the roles of the left- and right-handed helicities are interchanged. Apart from this interchange, the evolution in \cite{Tashiro:2012mf} is correct.  In the present analysis we have corrected both signs.}
\begin{eqnarray}
	\gbc &=& - \frac{\gamma_D}{2\pi^2} \left( 2 y_{L}^2 \gamma_{L} + 
6 y_{Q}^2 \gamma_{Q} - 3 y_{u_R}^2 \gamma_{u_R} - 3 y_{d_R}^2 \gamma_{d_R} - 
y_{e_R}^2 \gamma_{e_R} \right)~, \label{gammaB}   \\
	\goc &=& \frac{\gamma_D}{4\pi^2} \left( 2 y_{L} \gamma_{L}^2 + 
6 y_{Q} \gamma_{Q}^2 - 3 y_{u_R} \gamma_{u_R}^2 - 3 y_{d_R} \gamma_{d_R}^2 - 
y_{e_R} \gamma_{e_R}^2 \right)~, \label{gammaomega} 
\end{eqnarray}
where $\gamma_s \equiv \mu_s/T$ is the comoving chemical potential for particle species $s$.  
The evolution of the chiral coefficient $\gamma_{B}$ is governed by the anomaly equation, \eref{eq:anomalyequations}, which can be reexpressed as 
\begin{align}\label{mueq}
	\partial_{\eta} \gbc &= - \frac{ \gamma_D}{2 \pi^2} \frac{760}{9} \frac{\alpha_Y}{4 \pi} \partial_\eta h_{Yc} \com
\end{align}
where $h_{Yc} \equiv h_Y/T^3$ is the comoving helicity density.  
We have ignored chirality flipping due to Yukawa interactions since we are considering temperatures above $T_{\rm Yuk} \gtrsim 10^2 \TeV$, see \eref{eq:Tyuk}.  
As we discuss below, we will not require a kinetic equation for $\gamma_{\omega}$. 

Although the expressions in the previous paragraph are general, we will now specify to the case of a non-turbulent medium by setting ${\bm v}=0$. 
Turbulence will be treated separately in the following section.  
Equations~(\ref{mhdeq})~and~(\ref{mueq}) have been solved previously in Refs.~\cite{Joyce:1997uy,Boyarsky:2011uy,Tashiro:2012mf} for the case ${\bm v}=0$.  
It is convenient to decompose the magnetic field into Fourier modes of definite helicity, $B^\pm (\eta, k)$, where $k=|{\bf k}|$ is the dimensionless comoving wavenumber.  
The general solution is 
\begin{align}
	|B^\pm (\eta,k)| &= |B^\pm_0(k)| \exp (\gamma_D K_p^2 \eta ) \exp \left [ -\gamma_D \left ( k \mp K_p \right )^2 \eta \right ]~, \label{eq:Bsol}  \\
	\gamma_{B}(\eta) &= \gamma_{B}(0) + \frac{\gamma_D}{2 \pi^2} \frac{760}{9} \frac{\alpha_Y}{4 \pi} \Bigl( h_{Yc}(0) - h_{Yc}(\eta) \Bigr) \com  \label{eq:hYcsol} \\
	K_p &\equiv \frac{1}{2\eta}\int_{0}^{\eta} d\eta ~\frac{\gamma_B}{\gamma_D} \com \label{Kpdefn}
\end{align}
where $|B^\pm_0(k)|$ is the initial spectrum and $\gamma_{B}(0)$ is the initial chiral-magnetic coefficient. 
The helicity is evaluated from the magnetic field modes using 
\begin{equation}\label{helestimate}
	h_{Yc} = \int \frac{dk}{2\pi^2} k (|B^+|^2 - |B^-|^2)~.
\end{equation}
Thus we can identify $B^+$ as the amplitude of right-helical modes and $B^-$ as left-helical.

As seen by \eref{eq:Bsol}, the nonzero chiral asymmetry ensures that one of the two helical modes experiences an exponential amplification, 
\begin{align}\label{eq:exp_amplify}
	B(\eta,k) = B_0(k) \ {\rm exp}( \gamma_D K_p^2 \eta ) \com
\end{align}
over the range of scales $0 < k < k_{\rm max}$ with $k_{\rm max} \equiv 2 K_p$.  
However, there are three reasons why this would-be amplification does not have any significant impact on the evolution of the hypermagnetic field.  

First, using the expression for $K_p$, \eref{Kpdefn}, we see that the exponent in \eref{eq:exp_amplify} is proportional to $\gamma_B^2$.  
We can explicitly evaluate $\gamma_B$ and $\gamma_{\omega}$ using Eqs.~(\ref{gammaB}, \ref{gammaomega}), and the initial conditions in \eref{musolution} to find
\begin{equation}\label{eq:initial}
	\gbc (t_i) = 0
	\qquad {\rm and} \qquad
	\goc (t_i) = 0~.
\end{equation}
That is, when the hypercharge sphaleron is in equilibrium, \eref{eq:HS_chempot} is satisfied and enforces the vanishing of precisely the same combination of chemical potentials that appears in $\gamma_B$.  
Although the individual particle species in the medium possess chiral asymmetry, 
the net chiral asymmetry required to drive the chiral magnetic and vortical effects is absent.
If the hypercharge sphaleron were to go out of equilibrium, then a chiral asymmetry may be regenerated via the evolution of the helicity, see \eref{mueq}.  
In order to fully characterize the magnetic field evolution in the presence of a chiral asymmetry that may have arisen from leptogenesis, we will suppose that $\gamma_B / \gamma_D \sim \mu_0 / T$ in the discussion below.  

Second, we can estimate the scale $k_{\rm max} \sim K_p$ from \eref{Kpdefn} as $K_p \sim \gamma_B / \gamma_D \sim \mu_0 / T$ where $\mu_0$ is the typical scale of the chiral asymmetries.  
For the case of leptogenesis, $\mu_0$ is fixed by the observed baryon asymmetry via \eref{lepbar}, and we obtain $k_{\rm max} \sim \mu_0 / T \sim 10^{-8}$.  
This corresponds to a physical length scale $l_{\rm min} = 1 / (k_{\rm max} T) \sim 10^{8} T^{-1}$.  
For comparison, the correlation length of the magnetic field was given by \eref{eq:xi_initial} to be $\xi \sim 10^2 T^{-1}$.  
Since $\xi \ll l_{\rm min}$, the exponential amplification does not affect the modes which carry most of the power.  
The outcome could be different if the chiral asymmetry and/or the magnetic field correlation length were larger.  

Third, the time scale for this growth is on the order of $\eta_{\rm grow} \sim (\gamma_D K_p^2(\eta))^{-1}$ for the fastest growing mode, which has $k \sim K_p(\eta)$.  
Using the same numbers as in our estimates above, we obtain $\eta_{\rm grow} \sim \gamma_D / \gamma_B^2 \sim (\gamma_D \mu_0^2 / T^2)^{-1} \sim 10^{18}$.  
Moving back to physical units (see footnote \ref{fn:conventions}), this time scale corresponds to a temperature of $T \sim 1 \GeV$.  
At such a low temperature, the Yukawa interactions are in equilibrium, chirality flipping becomes active, and the chiral asymmetry can be erased. 
If the initial chiral asymmetry were larger, instead of being constrained by the baryon asymmetry, only then would chirality play a significant role in the evolution of the magnetic 
field \cite{Joyce:1997uy}.


\section{Magnetic Field Evolution in a Turbulent Medium}
\label{sec:turbulence}

In the previous section, we considered two scenarios in which the magnetic field evolves in a non-turbulent medium, and we found that the field dissipates rapidly.  
The outcome is more promising if the cosmological medium is turbulent, because the associated inverse cascade provides a means of transporting power from large to small scales.  
As a third and final case we will suppose that turbulence is present in the cosmological medium at the epoch of magnetogenesis.  
We will elaborate on the assumption of turbulence at the end of this section.  
To be clear, however, let us remark that leptogenesis is not typically accompanied by turbulence, and this should be understood as an additional assumption.  
Turbulence may arise, for example, as the result of a first order phase transition.  
In this section, we adapt the results of analytical and numerical studies \cite{Jedamzik:1996wp, Banerjee:2004df, Kahniashvili:2012uj, Saveliev:2013uva}.  
These show the development of an inverse cascade during which the comoving coherence scale of a helical magnetic field grows in proportion to $\eta^{2/3}$ and that the comoving magnetic helicity density is conserved.  

The inverse cascade scaling follows from the conservation of magnetic helicity, see \eg \, \cite{Banerjee:2004df}.  
From \eref{eq:hY_def} the helicity density is given by $h_Y \sim \xi \, \rho_{B_Y}$ where it is assumed that the spectrum peaks at a comoving length scale $\xi$ for which the comoving hypermagnetic energy density is $\rho_{B_Y}$.  
If $\rho_{B_Y} \sim \eta^{-n}$ then conservation of magnetic helicity requires $\xi \sim \eta^{n}$.  
The decay rate of magnetic energy density can be approximated as $d \rho_{B_Y} / d\eta \approx \rho_{B_Y} / \tau_{\rm eddy}$ where $\tau_{\rm eddy} = \xi / v_{\xi}$ is the eddy turnover time on the length scale $\xi$ and $v_{\xi}$ is the fluid velocity on that scale.  
In the turbulent regime, the fluid velocity $v \approx v_A$ where the Alfven velocity $v_A$ satisfies $v_A \sim B_Y / \sqrt{\rho_{\rm tot}}$.  
Using the above expressions the rate equation becomes $d \rho_{B_Y} / d\eta \sim \rho_{B_Y}^{5/2} / h_Y \sqrt{\rho_{\rm tot}}$.  
Conservation of magnetic helicity ensures that $h_Y$ is constant, and the comoving total energy density is constant as well.  
Then the rate equation is solved by $\rho_B \sim \eta^{-2/3}$ and therefore $\xi \sim \eta^{2/3}$.  

As we discussed in Sec.~\ref{sec:magnetogenesis}, we expect hypercharge sphalerons at a temperature $T_i \sim 10^{10} \GeV$ to produce magnetic fields on a length scale $\xi_i \sim (\alpha_Y T_i)^{-1} \sim 2 \times 10^{-22} \cm$.
This coherence scale grows due to inverse cascade and cosmological expansion.  
In a turbulent medium, the eddy turn-over time on this scale is $\delta t_{i} \sim \xi_i/v$ where $v = |{\bf v}|$, and we assume $v \sim 1$.  
To move to conformal coordinates we use $\delta \eta / \eta = \delta t / 2 t$, which is applicable in a radiation-dominated universe, and find that the eddy turn-over time corresponds to 
\begin{equation}
	\delta\eta_{i} \sim \frac{\eta_i \xi_i}{2 v t_i}~.
\end{equation}
The inverse cascade causes the coherence scale to grow by a factor of $(\eta / \delta \eta_{i})^{2/3}$ \cite{Saveliev:2013uva, Kahniashvili:2012uj}\footnote{More accurately the factor is $[(\eta - \eta_i + \delta \eta_i) / (\delta \eta_i) ]^{2/3}$, and for $\eta - \eta_i > {\rm few} \times \delta \eta_i$ the scaling in the text is obtained.  This expression generalizes the formula found in Refs.~\cite{Saveliev:2013uva, Kahniashvili:2012uj}, namely $(\eta / \eta_i)^{2/3}$, where the physical system of interest had a larger coherence scale, comparable to the Hubble scale, and correspondingly $\delta \eta_i \sim \eta_i$.  
}.  
Additionally, the coherence scale grows due to the cosmological expansion by a factor of $(a / a_i) = (T_i/T)[g_{\ast}(t_i) / g_{\ast}(t)]^{1/3}$ where $a(t)$ is the cosmological scale factor and $g_{\ast}(t)$ is the number of relativistic species at time $t$.  
At the time of leptogenesis $g_{\ast}(t_i) \approx 106.75$.  
Taken together, the coherence scale becomes 
\begin{equation}\label{xit}
	\xi(t) \approx \xi_i \frac{a}{a_i} \left ( \frac{\eta}{\delta\eta_{i}} \right )^{2/3}
	\approx \frac{[g_{\ast}(t_i)]^{1/6}}{[g_{\ast}(t)]^{1/2}} \left( \frac{45 v^2}{4 \pi^3 \alpha_Y} \right)^{1/3} \frac{1}{T} \left( \frac{M_P}{T} \right)^{2/3} \com
\end{equation}
where we have used $t = (2H)^{-1}$ with $H$ given by \eref{H}.  
It is interesting to note that $\xi(t)$ is independent of the initial temperature as a result of $\xi_{i} \propto T_{i}^{-1}$.
By the end of the radiation-dominated era, \ie, the epoch of radiation-matter equality, we have $T_{\rm eq} \approx 1~{\rm eV} \approx 5 \times 10^4~{\rm cm}^{-1}$ and $g_{\ast}(t_{\rm eq}) \approx 3$, and the coherence scale has grown to
\be
	\xi_{\rm eq} \sim 5 \times 10^{14}~ {\rm cm} \per
\ee
Subsequent evolution during the matter-dominated epoch until today when $T_0 \approx 2 \times 10^{-4} \eV$ yields an additional factor of $(a_{0} / a_{\rm eq}) \sim (T_{\rm eq} / T_0) \sim 3000$ growth due to cosmological expansion.  
The inverse cascade brings in a factor of $(\eta_0 / \eta_{\rm eq})^{2/3} \sim (T_{\rm eq}/T_0)^{1/3} \sim 20$, where $\eta_0$ and $\eta_{\rm eq}$ are the conformal times today and at radiation-matter equality.  
Therefore, the coherence scale at the present epoch is estimated to be
\be\label{eq:xi0_estimate}
	\xi_{0} \sim 2 \times 10^{19}~{\rm cm} \sim 10~{\rm pc} \per
\ee
If turbulence is not present ever since the time of leptogenesis or has damped away before the present day, contrary to our initial assumption, then \eref{eq:xi0_estimate} is an overestimate.  

For comparison, the coherence scale of the magnetic field created during electroweak baryogenesis \cite{Vachaspati:2001nb} can be larger because the dynamics of the first order phase transition can introduce additional correlations of the magnetic field on a scale set by the bubble radius, $R$ \cite{Ng:2010mt}.  
Depending on parameters, $R$ may be comparable to the horizon size at the electroweak epoch.  
If we take $\xi_i \sim H^{-1}$ at $T \sim 100 \GeV$, the estimate of \eref{xit} gives $\xi_0 \sim 1~{\rm Mpc}$.   

Once we have obtained the present coherence scale, the magnetic energy density can be estimated using Eq.~(\ref{hrhoB}) and the conservation of helicity, Eq.~(\ref{eq:hY_numerical}): 
\begin{equation}\label{rhoBt0}
\rho_B (t_0) \sim \frac{\alpha_Y^4}{\ecp~\alpha_{\rm w}^2} \, \frac{h_Y}{\xi_0}
            \sim  10^{-23} \left(\frac{10^{-8}}{\ecp}\right)~T_0^4~,
\end{equation}
where $T_{0} \approx 10^{-4}~{\rm eV}$ and we used $\xi_0 \approx 10^{20}~ T_{0}^{-1}$. Using $\rho_B = B^2/8 \pi$ gives the estimate
\begin{equation}\label{eq:B_today}
	B_0 \sim 10^{-18} \left(\frac{10^{-8}}{\ecp}\right)^{1/2} {\rm Gauss}~,
\end{equation}
where we used the relation $1~{\rm Gauss}  = 7 \times 10^{-20} \GeV^2$.  
For $\ecp \sim 10^{-13}$, the smallest value consistent with our calculations [see below \eref{rhoB} and the caveat below \eref{eq:ecp_numerical}], the field strength can be as large as $B_0 \sim 10^{-15}~{\rm G}$.  
Note that a magnetic field with this magnitude is well below the cosmic microwave background bound, $B \lesssim 10^{-9} \, {\rm G}$ \cite{Ade:2013zuv}.  
It is, however, remarkably close to the blazar bound, $B \gtrsim 10^{-16} \, {\rm G}$ \cite{Neronov:1900zz, Tavecchio:2010mk}, but the correlation length, given by \eref{eq:xi0_estimate}, is probably too small to expect any signal.  

A similar dimensional estimate can be made for the magnetic field generated during electroweak baryogenesis \cite{Vachaspati:2001nb}.  
The only difference is the factor of $\alpha_Y^4/\alpha_{\rm w}^2 \sim10^{-5}$ in \eref{rhoBt0} that arose from CP violation arguments.  
Since there is no competition between lepton/anti-lepton annihilations versus anomalous conversion of lepton to baryon number, this factor is absent for electroweak baryogenesis.  
This leads to an enhancement factor of $\sim10^5$ in the energy density, and $\sim 10^2$ in the magnetic field strength.  
This enhancement can be partially offset by the increased coherence scale in electroweak baryogenesis scenarios as discussed above.  

In the above discussion we have ignored effects of electroweak symmetry
breaking which will convert hypercharge magnetic fields into electromagnetic
magnetic fields.  
The field strengths are related by an $O(1)$ factor proportional to $\cos \theta_W$ where $\theta_W$ is the weak mixing angle \cite{Joyce:1997uy}.

Let us now elaborate on the assumption of turbulence.  
The criterion for turbulence is that the kinetic Reynolds number satisfies $R_e \gg 1$.  
On the length scale of interest, $\xi \sim (\alpha_Y T)^{-1} \sim 100 \, T^{-1}$, the Reynolds number is given by $R_e = \tau_{\rm diss} / \tau_{\rm eddy}$ where $\tau_{\rm eddy} = \xi / v_{\xi}$ is the eddy turnover time scale, $v_{\xi}$ is the velocity of the fluid on the scale $\xi$, and $\tau_{\rm diss}$ is the timescale for dissipation.  
Dissipation results from the finite mean free path of particles in the fluid.  
The right-chiral leptons interact only through the hypercharge gauge field, and therefore they have the largest mean free path (MFP) of any species in the plasma.  
Their MFP is estimated as $l_{\rm mfp}^{\ell_R} \sim (\alpha_Y^2 g_{\ast} T)^{-1} \sim 100 \, T^{-1}$ where the factor of $g_{\ast} \sim 100$ arises from the number density of hypercharge-carrying species (see, {\it e.g.}, \cite{Arnold:2000dr}). 
Since $\xi \sim l_{\rm mfp}^{\ell_R}$ the right-chiral leptons are effectively free-streaming on the length scale of the magnetic field, and their dissipative effect is manifest as a drag force $\dot{\bf v} + ({\bf v} \cdot {\bm \nabla}){\bf v} + \ldots =  - \alpha {\bf v}$ \cite{Jedamzik:1996wp}, where the ellipsis denotes additional terms due to magnetic fields.  
The drag coefficient is estimated as $\alpha \sim (g / g_{\ast}) (1 / l_{\rm mfp}^{\ell_R}) \sim 10^{-4}T$ where the suppression factor $(g/g_{\ast}) \sim 10^{-2}$ arises because the $g = 3$ species of right-chiral leptons are a subdominant component of the electroweak plasma \cite{Jedamzik:1996wp}.  
The associated dissipative time scale is $\tau_{\rm diss} = 1 / \alpha \sim 10^{4} T^{-1}$, and the Reynolds number is found to be $R_e(\xi) = v_{\xi} / (\alpha \xi) \sim 10^2 v_{\xi}$.  
Therefore the dissipative drag force due to right-chiral leptons is negligible provided that $v_{\xi} \gtrsim 10^{-2}$.  

The left-chiral leptons have the second largest MFP, $l_{\rm mfp}^{\ell_L} \sim (\alpha_W^2 g_{\ast} T)^{-1} \sim 10 \, T^{-1}$, since they carry weak isospin but not color.  
Since $\xi \gg l_{\rm mfp}^{\ell_L}$, the dissipative force is a diffusive force $\dot{\bf v} + ({\bf v} \cdot {\bm \nabla}){\bf v} + \ldots =  \eta \nabla^2 {\bf v}$. 
The viscosity is given by $\eta \sim (g / g_{\ast}) l_{\rm mfp}^{\ell_L} \sim 10^{-1} T^{-1}$, and the dissipative time scale is $\tau_{\rm diss} = \xi^2 / \eta \sim 10^{5} T^{-1}$.  
As before, we compare with the eddy turnover time scale and find the Reynold's number to be $R_e = v_{\xi} \xi / \eta \sim 10^{3} v_{\xi}$.  
Therefore the diffusive force is negligible provided that $v_{\xi} \gtrsim 10^{-3}$, and the system is turbulent.  

It is important to recognize the factor of $(g_{\ast} / g) \sim 100$ in these estimates.  
The system remains turbulent even though the magnetic field correlation length is comparable to the mean free path of right-chiral leptons, because the diffusing species represent a subdominant component of the electroweak plasma.  
The colored particles are the most abundant and the most tightly coupled ($l_{\rm mfp} \sim T^{-1}$).  

If the fluid velocity at the hypermagnetic field coherence length scale is small, \ie \ $v_{\xi} \lesssim 10^{-2}$, then the medium will not be turbulent and the hypermagnetic field will dissipate immediately.  
The bubble wall velocity at a first order phase transition depends on the properties of the transition (\eg, degree of supercooling) but $v_{\rm wall} = O(0.01 - 0.1)$ is typical \cite{Huet:1992ex}, and the fluid velocity is comparable \cite{Leitao:2010yw}.  
In the case of magnetic fields created at the electroweak or QCD epochs, it is known that the late time evolution consists of intermittent phases of turbulence and viscous damping (as the neutrinos and photons decouple) during which the inverse cascade is halted, but that the overall effect is consistent with the $\xi \sim \eta^{2/3}$ scaling \cite{Banerjee:2004df}.  
We expect that the situation will be similar for magnetogenesis at the leptogenesis epoch.  


\section{Summary and discussion}
\label{sec:disc}

We have found that leptogenesis produces {\it right-handed} helical magnetic fields.  
We have investigated the chiral-magnetic effect in a non-turbulent setting and concluded that it will not exponentially amplify the magnetic field.  
We have then discussed the evolution of the magnetic field assuming a turbulent medium.  
Borrowing results from earlier studies, including numerical simulations, we estimate the magnetic field coherence scale to be $\xi_0 \sim 10~{\rm pc}$ at the present epoch, an estimate that does not depend on the initial time of leptogenesis.  
The field strength at the present epoch is estimated using helicity conservation, and can be as large as $B_0 \sim 10^{-15} ~\sqrt{10^{-13}/\ecp}~{\rm Gauss}$. 
In the case of hierarchical right-handed neutrinos, $\ecp$ is bounded from above by the so called Davidson-Ibarra bound: $\ecp \lesssim 10^{-6} (M_{N_R}/10^{10}~{\rm GeV})$  \cite{Davidson:2002qv}.  
We note that Eq.~(\ref{rhoB}) breaks down for values $\ecp \lesssim 10^{-13}$, where we cannot ignore the backreaction of the magnetic field on the plasma anymore.  
This gives a range of values $10^{-13} \lesssim \ecp \lesssim 10^{-6}$, where our estimates apply.  
Using \eref{eq:B_today}, we find that in the standard leptogenesis scenario accompanied by turbulence afterwards leads to the present magnetic field strength in the range $10^{-18} \lesssim B_{0}/{\rm Gauss}~ \lesssim 10^{-15}$.

The {\em right-handed} helicity of the magnetic field from leptogenesis can be understood
heuristically by noting that
\begin{equation}
\Delta h_A \sim \Delta h_Y \sim - \Delta n_{e_R}
\end{equation}
and the equilibration process described in Sec.~\ref{sec:equil}
gives $\Delta n_{e_R} < 0$. Since the initial electromagnetic
magnetic field vanishes, we obtain $h_{A} > 0$.

In contrast to leptogenesis, it is known that magnetogenesis during electroweak baryogenesis yields a {\it left-helical} magnetic field \cite{Vachaspati:2001nb}. 
Measurements of the primordial magnetic field can provide an additional handle on different baryogenesis scenarios, and as such we review here the heuristic arguments behind the left-helical magnetic fields in electroweak baryogenesis.  
Consider an electroweak process that leads to production of baryon number. 
For convenience, we turn off all gauge fields except  for the $Z$ gauge field.  
Then the baryon number anomaly equation, (\ref{baryonanomaly}), gives 
\begin{equation}\label{baryonZonly}
\Delta n_{\rm bar} = \frac{N_g (g^2+g^{\prime 2})}{32\pi^2}\cos(2\theta_W) \, \Delta 
\int_V \frac{d^3x}{V} ~
\vcz\cdot \vcb_Z ~
\propto ~ \Delta h_Z \com
\end{equation}
where, in the unitary gauge, the $Z_\mu$ and $A_\mu$ (electromagnetic) gauge fields are 
defined in terms of the weak and hypercharge gauge fields by
\ba
Z_\mu &=& \cos\theta_W W_\mu^3 - \sin\theta_W Y_\mu~, \label{Zdefn}
\\
A_\mu &=& \sin\theta_W W_\mu^3 + \cos\theta_W Y_\mu~, 
\ea
where $\theta_W=\tan^{-1}(g'/g)$ is the weak mixing angle of the Standard Model.
For $\Delta n_{\rm bar} > 0$, we require processes for which $\Delta h_Z > 0$.  
Now consider the decay of a sphaleron-like complex of $Z$ field.  
At the initial moment, the configuration carries $Z-$helicity and we take $h_Z(t_i) \ne 0$. 
In the final stage, all $Z$ fields must have decayed and the final $Z-$helicity vanishes.
Therefore the change in the $Z-$helicity is positive only if the initial $Z-$helicity is negative: $h_{Z_{i}} < 0$. 
From Eq.~(\ref{Zdefn}), we find that the hypercharge magnetic field in the initial configuration must also have carried negative helicity: $h_{Yi} < 0$.
After the sphaleron-like complex of $Z$ field has decayed, electromagnetic magnetic fields are produced. 
This can be thought of as the conversion of hypercharge magnetic fields into electromagnetic magnetic fields.  
Thus the final magnetic field that is produced also carries negative helicity, \ie, the magnetic field produced during baryogenesis is left-handed.

Extending our comparison beyond electroweak baryogenesis, we may ask which other baryogenesis mechanisms can give rise to helical magnetic fields.  
In analogy with the leptogenesis case, we expect that the models for which nonperturbative gauge field configurations play a central role in the creation of the baryon asymmetry are the most compelling candidates for magnetogenesis.  
These include, for example, some among a relatively new class of models in which sphalerons serve to connect the Standard Model to the dark sector  \cite{Buckley:2010ui, Haba:2010bm, Shelton:2010ta}.  
On the other hand, models which do not require nonperturbative B-number violation, such as Affleck-Dine baryogenesis \cite{Affleck:1984fy}, may nevertheless be worthwhile to investigate from a magnetogenesis perspective since the ubiquitous sphalerons will still process the $B+L$ asymmetry and may, thereby, generate a magnetic field.  

As we discussed above, chiral effects play no role in the evolution of the magnetic field when the chiral asymmetry is comparable in magnitude to the observed baryon asymmetry of the universe, which is tiny.  
This behavior, which has been noted previously \cite{Joyce:1997uy}, follows from the fact that the timescales of the chiral magnetic and vortical effects are inversely proportional to the magnitude of the chiral asymmetry.  
It is interesting to consider how our analysis would change if leptogenesis gives rise to a large initial lepton asymmetry, $\mu_0 / T = O(1)$\footnote{Interestingly, if the lepton asymmetry were very large, $\mu_0 / T \gtrsim 1$, then electroweak symmetry restoration may not occur \cite{Linde:1976kh}, in which case the required baryon asymmetry could be produced from an initial large lepton asymmetry since the electroweak sphalerons would be suppressed \cite{Liu:1993am,Casas:1997gx}. Note, however, that the lepton asymmetry in neutrinos is bounded as $|\mu_0/T| \lesssim 10^{-2}$ due to its effect the primordial element abundances and the CMB anisotropy (see, e.g., Ref.~\cite{Schwarz:2012yw} for recent limits), and therefore significant dilution of the lepton asymmetry is required between the leptogenesis era and BBN in such a scenario.}, and subsequently some unspecified physics leads to a suppression of the baryon asymmetry down to the observed level.  
A large injection of entropy, for example, could provided the desired suppression.  
In this case, the chiral magnetic effect would lead to an exponential growth of the helical magnetic field \cite{Joyce:1997uy,Boyarsky:2011uy,Tashiro:2012mf}.  
The resultant field amplitude today could be much larger than our estimate in \eref{eq:B_today}, but our estimates for the helicity, \eref{eq:hY_numerical}, and coherence length, \eref{eq:xi0_estimate}, would not be significantly affected by the presence of chiral effects.  

As we have seen, leptogenesis leaves its signature in the form of relic helical magnetic fields.  
Although measurements of the properties of these fields are, at present, extremely challenging, such observations would undoubtedly be viewed as a great boon. Therefore, despite the practical difficulties, it is nevertheless important to explore the extent to which such a boon would reveal the nature of the very early universe. Measuring the sign of the relic magnetic field's helicity could, for example, help to determine how the baryon asymmetry of the universe arose.  
This is particularly important in the case of leptogenesis since it is otherwise difficult to directly test the high scale physics that is involved.  
In fact, recently a technique has been proposed to allow for a measurement of the helicity of the intergalactic magnetic field using gamma ray data from blazars \cite{Tashiro:2013bxa} and the diffuse gamma ray data \cite{Tashiro:2013ita}.  
Future measurements of relic magnetic fields will provide unique insight into early universe cosmology and physics beyond the Standard Model, may allow discrimination between different matter genesis scenarios, and could help to distinguish cosmological versus astrophysical production mechanisms of magnetic fields.


\acknowledgments
TV thanks Kfir Blum, Juan Maldacena, Oleg Ruchayskiy, Hiroyuki Tashiro, and Edward Witten for discussion, and
is grateful to the Institute for Advanced Study, Princeton, for hospitality.  
AL thanks Mt. Stromlo Observatory (Australia National University) for hospitality during the completion of this work.  
This work was supported by the DOE at ASU. 


\appendix

\section{Signs and conventions}
\label{sec:signs}

In this Appendix we use the Minkowski metric in Cartesian coordinates: 
$\eta_{\mu\nu}={\rm diag}(+1, -1, -1, -1)$.  
The contravariant vector potential is written as $A^{\mu} = (\phi , \vca)$. The
covariant vector is obtained by contraction with the 4-dimensional Minkowski metric,  
$A_{\mu} = \eta_{\mu\nu} A^\nu = (\phi, -{\bf A})$, and the contravariant and covariant
forms of the 3-vector, $\vca$, are related by contraction with the 3-dimensional Euclidean
metric $\eta^{(3)}_{ij}=\delta_{ij}$. Thus $\vca^i=\vca_i$. 

The field strength tensor is defined by $F_{\mu\nu} \equiv \partial_\mu A_\nu - \partial_\nu A_\mu$ where $\partial_{0} = \partial / \partial t = \partial_{t}$ and $\partial_{i} = \partial / \partial x^i = {\bm \nabla}_i$.  
The electric and magnetic field three vectors are defined as
\begin{equation}
	\vce_i = F_{0i} = -\partial_t \vca_i - {\bm \nabla}_i \phi 
	\qquad {\rm and} \qquad
	\vcb_i = - \frac{1}{2} \epsilon^{0ijk} F_{jk} = ({\bm \nabla}\times \vca )_i~,
\end{equation}
where $\epsilon^{\mu \nu \alpha \beta}$ is the totally antisymmetric rank 4 tensor with the normalization $\epsilon^{0123}=+1$.  
A sum over repeated indices is understood. 
The dual field strength tensor is defined by ${\tilde F}^{\mu\nu} \equiv \frac{1}{2} \epsilon^{\mu\nu\alpha\beta} F_{\alpha\beta}$.  
It is straightforward to show that 
\begin{equation}
	F_{\mu\nu} {\tilde F}^{\mu\nu} 
	= 2 \epsilon^{0ijk} F_{0i} F_{jk}
	= - 4 \vce \cdot \vcb \com
\end{equation}
where $\vce \cdot \vcb = \vce_i \vcb_i$.  

We define the spatially averaged helicity density by, 
\begin{equation}
	h(t) \equiv \lim_{V \to \infty} \frac{1}{V} \int_V d^3x ~ \vca \cdot \vcb \per
\end{equation}
The rate of change is
\begin{equation}
	\partial_t h 
	= - \lim_{V \to \infty} \frac{2}{V} \int_V d^3x ~ \vce \cdot \vcb 
	= + \lim_{V \to \infty} \frac{1}{2V} \int_V d^3 x ~ F_{\mu\nu} {\tilde F}^{\mu\nu}~,
\end{equation}
where we have used the source free Maxwell's equations, $\partial_{t} \vcb = - {\bm \nabla} \times \vce$ and ${\bm \nabla} \cdot \vcb = 0$, and dropped the surface terms.  
Finally the integral over time gives the change $\Delta h = h(\infty) - h(-\infty)$ to be
\begin{equation}\label{eq:Deltah}
	\Delta h 
	= - \lim_{V \to \infty} \frac{2}{V} \int dt \int_V d^3x ~ \vce \cdot \vcb 
	= + \lim_{V \to \infty} \frac{1}{2V} \int dt \int_V d^3 x ~ F_{\mu\nu} {\tilde F}^{\mu\nu}~.
\end{equation}


\section{Anomaly equations}
\label{sec:anomeqs}

This appendix summarizes the chiral anomaly equations.  
In order to set our notation, we will first consider the chiral anomaly in massless quantum electrodynamics, and we then proceed to the Standard Model.  
Consider the Abelian gauge theory defined by 
\begin{align}
	\mathcal{L} = i \overline{\Psi} \gamma^{\mu} \left( \partial_{\mu} - i e A_{\mu} \right) \Psi - \frac{1}{4} F_{\mu \nu} F^{\mu \nu} \per
\end{align}
This Lagrangian is invariant under the transformations
\begin{align}
	&\text{Local Gauge Transformation:} \qquad \left\{ 
	\begin{array}{l}
	\Psi(x) \to e^{i \theta(x)} \Psi(x) \\
	A_{\mu}(x) \to A_{\mu}(x) - \frac{1}{e} \partial_{\mu} \theta(x)
	\end{array}
	\right. \\
	&\text{Global Chiral Transformation:} \qquad \left\{ \Psi(x) \to e^{i \gamma^5 \alpha} \Psi(x) \right. 
\end{align}
where $\gamma^{5} \equiv - \frac{i}{4!} \epsilon_{\mu \nu \rho \sigma} \gamma^{\mu} \gamma^{\nu} \gamma^{\rho} \gamma^{\sigma} = i \gamma^{0} \gamma^{1} \gamma^{2} \gamma^{3}$ (recall that $\epsilon^{0123} = - \epsilon_{0123} = +1$).  
The corresponding classically conserved currents are 
\begin{align}\label{eq:example_currents}
	j^{\mu} \equiv \overline{\Psi} \gamma^{\mu} \Psi
	\qquad {\rm and} \qquad
	j_{5}^{\mu} \equiv \overline{\Psi} \gamma^{\mu} \gamma^{5} \Psi \per
\end{align}
At the quantum level, the electromagnetic current is still conserved, but the Abelian anomaly induces a nonzero divergence for the chiral current \cite{Adler:1969gk, Bell:1969ts}:  
\begin{align}\label{eq:example_anomaly_eqns}
	\partial_{\mu} j^{\mu} & = 0 \\
	\partial_{\mu} j^{\mu}_{5} & = - 2 \frac{e^2}{16 \pi^2} F_{\mu \nu} \tilde{F}^{\mu \nu}
\end{align}
where $\tilde{F}^{\mu \nu} \equiv \frac{1}{2} \epsilon^{\mu \nu \rho \sigma} F_{\rho \sigma}$.  
It is convenient to move the the two-component spinor notation by writing
\begin{align}
	\Psi = \begin{pmatrix} (\psi_L)_{\alpha} \\ (\psi_R)^{\dot{\alpha}} \end{pmatrix}
	\quad , \quad
	\gamma^{\mu} = \begin{pmatrix} 0 & \sigma^{\mu} \\ \bar{\sigma}^{\mu} & 0 \end{pmatrix} 
	\quad , \quad {\rm and} \quad
	\gamma^{5} = \begin{pmatrix} -1 & 0 \\ 0 & 1 \end{pmatrix}
\end{align}
where $\sigma^\mu= (1,{\vec \sigma})$ and ${\bar \sigma}^\mu = (1, -{\vec \sigma})$ and where $\sigma^i$ are the Pauli spin matrices.  
The spinors $\psi_L$ and $\psi_R$ have the same fermion number and electromagnetic charge as $\Psi$, but they have left- and right-handed chirality, respectively.  
In this notation the currents, \eref{eq:example_currents}, become
\begin{align}
	j^{\mu} & = + \psi_L^{\dagger} \bar{\sigma}^{\mu} \psi_L + \psi_R^{\dagger} \sigma^{\mu} \psi_R \\
	j_{5}^{\mu} & = - \psi_L^{\dagger} \bar{\sigma}^{\mu} \psi_L + \psi_R^{\dagger} \sigma^{\mu} \psi_R \com
\end{align}
and divergence equations, \eref{eq:example_anomaly_eqns}, can be written as 
\begin{align}\label{eq:example_chiral_anom}
	\partial_{\mu} j_L^{\mu} & = + \frac{e^2}{16 \pi^2} F_{\mu \nu} \tilde{F}^{\mu \nu} \nonumber \\
	\partial_{\mu} j_R^{\mu} & = - \frac{e^2}{16 \pi^2} F_{\mu \nu} \tilde{F}^{\mu \nu} 
\end{align}
where
\begin{align}
	j_{L}^{\mu} \equiv \psi_L^{\dagger} \bar{\sigma}^{\mu} \psi_L 
	\qquad {\rm and} \qquad 
	j_{R}^{\mu} \equiv \psi_R^{\dagger} \sigma^{\mu} \psi_R \per
\end{align}
The signs in \eref{eq:example_chiral_anom} are particularly important.  
The spin indices can be written explicitly as $j_L^{\mu} = (\psi_L^{\dagger})_{\dot{\alpha}} (\bar{\sigma}^{\mu})^{\dot{\alpha} \alpha} (\psi_L)_{\alpha}$ and $j_R^{\mu} = (\psi_R^{\dagger})^{\alpha} (\sigma^{\mu})_{\alpha \dot{\alpha}} (\psi_R)^{\dot{\alpha}}$.

The Standard Model currents are defined by
\begin{align}
\begin{array}{l}
	j^{\mu}_{Q^i} \equiv (Q)^{i \, \dagger} \bar{\sigma}^{\mu} (Q)^i \\
	j^{\mu}_{u_R^i} \equiv (u_R)^{i \, \dagger} \sigma^{\mu} (u_R)^{i} \\
	j^{\mu}_{d_R^i} \equiv (d_R)^{i \, \dagger} \sigma^{\mu} (d_R)^{i} \\
	j^{\mu}_{L^i} \equiv (L)^{i \, \dagger} \bar{\sigma}^{\mu} (L)^i \\
	j^{\mu}_{e_R^i} \equiv (e_R)^{i \, \dagger} \sigma^{\mu} (e_R)^{i} 
\end{array} \com
\end{align}
where suppressed gauge group indices are summed, but the generation index $i$ is not summed.  
Each current is anomalous by virtue of gauge interactions: 
\begin{align}\label{eq:anomalyequations}
\begin{array}{llllll}
	\partial_{\mu} j^{\mu}_{Q^i} = 
	&+\frac{1}{2}(N_w) \frac{g_s^2}{16 \pi^2} G_{\mu \nu}^A \tilde{G}^{A \, \mu \nu} 
	&+ 
	&\frac{1}{2} (N_c) \frac{g^2}{16 \pi^2} W_{\mu \nu}^a \tilde{W}^{a \, \mu \nu} 
	&+ 
	&\frac{1}{4} (N_c N_w y_Q^2) \frac{g^{\prime \, 2}}{16 \pi^2} Y_{\mu \nu} \tilde{Y}^{\mu \nu} \\
	\partial_{\mu} j^{\mu}_{u_R^i} = 
	&- \frac{1}{2} \frac{g_s^2}{16 \pi^2} G_{\mu \nu}^A \tilde{G}^{A \, \mu \nu} 
	& 
	&
	&- 
	&\frac{1}{4} (N_c y_{u_R}^2) \frac{g^{\prime \, 2}}{16 \pi^2} Y_{\mu \nu} \tilde{Y}^{\mu \nu} \\
	\partial_{\mu} j^{\mu}_{d_R^i} = 
	&- \frac{1}{2} \frac{g_s^2}{16 \pi^2} G_{\mu \nu}^A \tilde{G}^{A \, \mu \nu} 
	&
	&
	&- 
	&\frac{1}{4} (N_c y_{d_R}^2) \frac{g^{\prime \, 2}}{16 \pi^2} Y_{\mu \nu} \tilde{Y}^{\mu \nu} \\
	\partial_{\mu} j^{\mu}_{L^i} = 
	&
	&
	&\frac{1}{2} \frac{g^2}{16 \pi^2} W_{\mu \nu}^a \tilde{W}^{a \, \mu \nu} 
	&+ 
	&\frac{1}{4} (N_w y_{L}^2) \frac{g^{\prime \, 2}}{16 \pi^2} Y_{\mu \nu} \tilde{Y}^{\mu \nu} \\
	\partial_{\mu} j^{\mu}_{e_R^i} = 
	&
	& 
	&
	&- 
	& \frac{1}{4} (y_{e_R}^2) \frac{g^{\prime \, 2}}{16 \pi^2} Y_{\mu \nu} \tilde{Y}^{\mu \nu} 
\end{array} \com
\end{align}
where $N_c = 3$ and $N_w = 2$ are the ranks of the non-Abelian gauge groups, and the hypercharges are
\begin{align}
	y_{Q} = \frac{1}{3} \com \quad 
	y_{u_R} = \frac{4}{3} \com \quad
	y_{d_R} = - \frac{2}{3} \com \quad
	y_{L} = - 1 \com \quad
	y_{e_R} = -2 \per
\end{align}
The tensors $G_{\mu \nu}^{A}$, $W_{\mu \nu}^{a}$, and $Y_{\mu \nu}$ are the ${\rm SU}(3)_c$, ${\rm SU}(2)_L$, and ${\rm U}(1)_Y$ field strength tensors, respectively.  
The dual tensor is denoted by a tilde, such as $\tilde{X}^{\mu \nu} \equiv \frac{1}{2} \epsilon^{\mu \nu \rho \sigma} X_{\rho \sigma}$.  
We use the convention $\epsilon^{0123} = +1$.  
We have only written the anomalous contribution to the various currents and dropped the tree-level mixing from Dirac mass terms.  

The terms on the right-hand side of \eref{eq:anomalyequations} can be related to the Chern-Simons numbers:
\begin{align}
	N_{\rm CS}^{(c)}(t) \equiv \frac{1}{2} \frac{g_s^2}{16 \pi^2} \int d^3 x \, k_s^0
	&\qquad {\rm with} \qquad&
	&k_s^{\mu} \equiv \epsilon^{\mu \alpha \beta \gamma} \left( G_{\alpha \beta}^{A} G_{\gamma}^{A} - \frac{g_s}{3} f_{ABC} G_{\alpha}^{A} G_{\beta}^{B} G_{\gamma}^{C} \right) \\
	N_{\rm CS}^{(L)}(t) \equiv \frac{1}{2} \frac{g^2}{16 \pi^2} \int d^3 x \, k_L^0
	&\qquad {\rm with} \qquad&
	&k_L^{\mu} \equiv \epsilon^{\mu \alpha \beta \gamma} \left( W_{\alpha \beta}^{a} W_{\gamma}^{a} - \frac{g}{3} \epsilon_{abc} W_{\alpha}^{a} W_{\beta}^{b} W_{\gamma}^{c} \right) \\
	N_{\rm CS}^{(Y)}(t) \equiv \frac{1}{2} \frac{g^{\prime \, 2}}{16 \pi^2} \int d^3 x \, k_Y^0
	&\qquad {\rm with} \qquad& 
	&k_Y^{\mu} \equiv \epsilon^{\mu \alpha \beta \gamma} \left( Y_{\alpha \beta} Y_{\gamma} \right) \per
\end{align}
One can verify that 
\begin{align}
	\partial_{\mu} k_s^{\mu} = G_{\mu \nu}^{A} \tilde{G}^{A \, \mu \nu}
	\quad , \quad
	\partial_{\mu} k_L^{\mu} = W_{\mu \nu}^{a} \tilde{W}^{a \, \mu \nu}
	\quad \, , \ {\rm and} \quad 
	\partial_{\mu} k_Y^{\mu} = Y_{\mu \nu} \tilde{Y}^{\mu \nu} \com
\end{align}
and therefore
\begin{align}
	\Delta N_{\rm CS}^{(c)} & = \frac{1}{2} \frac{g_s^2}{16 \pi^2} \int d^4 x \, G_{\mu \nu}^{A} \tilde{G}^{A \, \mu \nu} \\
	\Delta N_{\rm CS}^{(L)} & = \frac{1}{2} \frac{g^2}{16 \pi^2} \int d^4 x \, W_{\mu \nu}^{a} \tilde{W}^{a \, \mu \nu} \\
	\Delta N_{\rm CS}^{(Y)} & = \frac{1}{2} \frac{g^{\prime \, 2}}{16 \pi^2} \int d^4 x \, Y_{\mu \nu} \tilde{Y}^{\mu \nu} 
\end{align}
where $\Delta N_{\rm CS} \equiv N_{\rm CS}(t = + \infty) - N_{\rm CS}(t = - \infty)$.  
The change in the Chern-Simons number, $\Delta N_{\rm CS}$, is gauge invariant eventhough $N_{\rm CS}(t)$ is not.  

The baryon and lepton number currents are defined as 
\begin{align}
\begin{array}{l}
	j^{\mu}_{\rm bar} \equiv \frac{1}{N_c} \sum_{i} \left( j^{\mu}_{Q^i} + j^{\mu}_{u_R^i} + j^{\mu}_{d_R^i} \right) \\
	j^{\mu}_{\rm lep} \equiv \sum_{i} \left( j^{\mu}_{L^i} + j^{\mu}_{e_R^i}  \right) 
\end{array} \per
\end{align}
Using \eref{eq:anomalyequations} the divergences of these currents are 
\begin{align}\label{baryonanomaly}
	\partial_{\mu} \, j^{\mu}_{\rm bar} = \partial_{\mu} \, j^{\mu}_{\rm lep} = \frac{N_g}{2} \left( \frac{g^2}{16 \pi^2} W_{\mu \nu}^a \tilde{W}^{a \, \mu \nu} - \frac{g^{\prime \, 2}}{16 \pi^2} Y_{\mu \nu} \tilde{Y}^{\mu \nu} \right)~,
\end{align}
where $N_g = 3$ is the number of generations.
Furthermore, one can easily verify that the hypercharge current
\begin{align}
	j_{Y}^{\mu} \equiv \sum_i ( y_Q j_{Q^i}^{\mu} + y_{u_R} j_{u_R^i}^{\mu} + y_{d_R} j_{d_R^i}^{\mu} + y_{L} j_{L^i}^{\mu} + y_{e_R} j_{e_R^i}^{\mu} )~,
\end{align}
is not anomalous $\partial_{\mu} j_{Y}^{\mu} = 0$.  



\providecommand{\href}[2]{#2}\begingroup\raggedright\endgroup

\end{document}